\documentclass[aps,prd,superscriptaddress,twoside,twocolumn,nofootinbib,showpacs,floatfix]{revtex4-1}
\usepackage{amsmath,amssymb}
\usepackage{graphicx,bm,braket}
\usepackage{subfigure}
\pdfpagewidth=\paperwidth
\pdfpageheight=\paperheight
\allowdisplaybreaks
\usepackage{epstopdf}
\newcommand*{\slashed}[1]{{#1\!\!\!/}}
\newcommand*{\hc}{\text{H.\,c.}}

\begin{document}

\title{\boldmath Nucleon and $\Delta$ resonances in $\gamma n \to K^+\Sigma^-$ photoproduction}

\author{Neng-Chang Wei}
\affiliation{School of Nuclear Science and Technology, University of Chinese Academy of Sciences, Beijing 101408, China}

\author{Ai-Chao Wang}
\affiliation{School of Nuclear Science and Technology, University of Chinese Academy of Sciences, Beijing 101408, China}

\author{Fei Huang}
\email{Corresponding author. Email: huangfei@ucas.ac.cn}
\affiliation{School of Nuclear Science and Technology, University of Chinese Academy of Sciences, Beijing 101408, China}

\author{Kanzo Nakayama}
\affiliation{Department of Physics and Astronomy, University of Georgia, Athens, Georgia 30602, USA}

\date{\today}

\begin{abstract}
The most recent data on beam asymmetries, $\Sigma$, and beam-target asymmetries, $E$, from the CLAS Collaboration, together with the previous data on differential cross sections and beam asymmetries from the CLAS and LEPS Collaborations, for the $\gamma n \to K^+\Sigma^-$ reaction are studied based on an effective Lagrangian approach in the tree-level Born approximation. The $t$-channel $K$ and $K^\ast(892)$ exchanges, the $u$-channel $\Sigma$ exchange, the interaction current, and the exchanges of $N$, $\Delta$, and their excited states in the $s$ channel are considered in constructing the reaction amplitudes to describe the available experimental data. The reaction mechanisms of $\gamma n \to K^+\Sigma^-$ are analyzed, and the associated resonances' parameters are extracted. The numerical results show that the $\Delta$ exchange and the $N(1710)1/2^+$, $N(1880)1/2^+$, $N(1900)3/2^+$, and $\Delta(1920)3/2^+$ resonance exchanges in the $s$-channel dominate the $\gamma n \to K^+\Sigma^-$ reaction in the lower energy region, and the $t$-channel $K^\ast(892)$ exchange plays a crucial role at forward angles in the higher energy region. 
\end{abstract}



\maketitle

\section{Introduction}   \label{Sec:intro}

The study of the properties of excited baryons such as their masses, widths, and transition form factors provides essential information in help understand the nonperturbative regime of quantum chromodynamics (QCD) and, in particular, the internal structure of baryons. Today, however, it is still a challenge to match the observed baryon spectrum with predictions from QCD-inspired phenomenological models \cite{Isgur:1977ef, Koniuk:1979vy} and/or lattice QCD calculations \cite{Edwards:2013, Lang:2017, Kiratidis:2017, Andersen:2018}, since, theoretically, more resonances are predicted than what have been found in experiments as listed in the most recent Review of Particle Physics (RPP) \cite{Zyla:2020zbs}. This is known as the \textit{missing resonance problem} in baryon spectroscopy.

Most of our earlier knowledge of baryon excitations have been gained from $\pi N$ scattering experiments or single-pion photoproduction experiments. As such, those ``missing resonances" might have been escaped from detections due to their weak couplings to the $\pi N$ channel. In this regard, concerted efforts are being made at electron accelerator facilities worldwide -- such as the Jlab, MAMI, ELSA, and Spring8 -- to measure the production reactions of mesons other than the pion. In this respect, in the past few years, the $KY$ ($Y=\Sigma,\Lambda$) photoproduction reaction has been receiving increasing attention, both experimentally and theoretically. 

Experimentally, as has been reviewed in Refs.~\cite{Mart:2019fau,Ireland:2019uwn}, a large number of data on differential cross sections and polarization observables have been measured in the past 20 years for the four elementary $K\Sigma$ photoproduction reactions, namely, $\gamma p \to K^0 \Sigma^+$, $\gamma p \to K^+ \Sigma^0$, $\gamma n \to K^0 \Sigma^0$, and $\gamma n \to K^+ \Sigma^-$ \cite{Glander:2003jw, McNabb:2003nf, Zegers:2003ux, Sumihama:2005er, Kohri:2006yx, Bradford:2005pt, Lleres:2007tx, Bradford:2006ba, Dey:2010hh, Jude:2013jzs, Paterson:2016vmc, Lawall:2005np, Castelijns:2007qt, Aguar-Bartolome:2013fqw, Akondi:2018shh, AnefalosPereira:2009zw}. A large body of these data have been measured in the proton channel, while the measurements of the neutron channel are comparatively scarce. Measurements of $KY$ photoproduction off neutron can offer additional constraints to learn more about the underlying reaction dynamics and, in particular, to determine the $N^\ast$ and $\Delta$ resonances that are excited in this reaction. Also, there are studies of Kaon photoproduction reactions involving excited hyperon in the final states, both experimentally \cite{Kohri:2009xe,Wieland:2010cq,Moriya:2013hwg,CLAS:2021osv} and theoretically \cite{Oh:2007jd,Wang:2014jxb,Wang:2016dtb,He:2013ksa,Wang:2020mdn,Wei:2021qnc,Zhang:2021iez,Kim:2017nxg,Yu:2017kng}. 

Very recently, the CLAS Collaboration has just released the data on beam asymmetries $\Sigma$ \cite{CLAS-beam} and beam-target asymmetries $E$ \cite{Zachariou:2020kkb} for the $\gamma n \to K^+\Sigma^-$ reaction. Their measurements of $E$ are the world's first measurement of this observable for $\gamma n \to K^+\Sigma^-$, and their measurements of $\Sigma$, compared with the very limited previous data from the LEPS Collaboration \cite{Kohri:2006yx}, cover broader ranges of kaon emission angle and photon energy. These new data on $\Sigma$ and $E$ are expected to help provide further constraints in the determination of the reaction dynamics in the $\gamma n \to K^+\Sigma^-$ reaction.

Theoretically, several works \cite{Vancraeyveld:2009qt,Mart:2019fau,Clymton:2021wof,Luthfiyah:2021yqe} have already been devoted to analyze the data for $K^+\Sigma^-$ photoproduction published before 2010 in Refs.~\cite{Kohri:2006yx,AnefalosPereira:2009zw}. In Ref.~\cite{Vancraeyveld:2009qt}, a Regge-plus-resonance model was adapted to analyze the LEPS data \cite{Kohri:2006yx} with amplitudes being constructed by $K$ and $K^{*}$ Regge-trajectory exchanges supplemented with a selection of $s$-channel resonance diagrams. In a series works of Refs.~\cite{Mart:2019fau,Clymton:2021wof,Luthfiyah:2021yqe}, the data from Refs.~\cite{Kohri:2006yx,AnefalosPereira:2009zw} for $K^+\Sigma^-$ photoproduction were analyzed together with the data for $K^0\Sigma^0$, $K^+\Sigma^0$, $K^0\Sigma^+$ photoproduction reactions. In Ref.~\cite{Mart:2019fau}, an isobar model was constructed from the Feynman diagrammatic approach for the background terms and multipoles formulation for the resonance terms to analyze the data, and all the resonances rated with two or more stars in RPP with spins ranging from 1/2 to 13/2 were included in the fit. In Ref.~\cite{Clymton:2021wof}, an updated isobar model with resonance terms also being constructed from Feynman diagrammatic approach was developed for $K\Sigma$ photoproduction. In Ref.~\cite{Luthfiyah:2021yqe}, the effects of the nucleon and $\Delta$ resonances with spin 11/2, 13/2, and 15/2 in $K\Lambda$ and $K\Sigma$ photoproduction reactions were investigated.

Stimulated by the recently released data on $\Sigma$ \cite{CLAS-beam} and $E$ \cite{Zachariou:2020kkb}, in the present work we perform an analysis of all the available data for $\gamma n \to K^+\Sigma^-$ \cite{Kohri:2006yx, AnefalosPereira:2009zw, CLAS-beam, Zachariou:2020kkb} based on an effective Lagrangian approach in the Bonn approximation. We consider the $t$-channel $K$ and $K^\ast(892)$ exchanges, the $u$-channel $\Sigma$ exchange, the interaction current, and the $s$-channel $N$, $\Delta$, and their excited states ($N^\ast$s and $\Delta^\ast$s) exchanges in constructing the reaction amplitude. It is shown that all the available data for $\gamma n \to K^+\Sigma^-$ can be well reproduced in our model. The reaction mechanisms of $\gamma n \to K^+\Sigma^-$ are analyzed and the parameters of the relevant $N$ and $\Delta$ resonances are extracted and compared with those quoted by RPP \cite{Zyla:2020zbs}.

We mention that, in principle, a coupled-channels approach that properly accounts for the unitarity and analyticity of the reaction amplitude should be employed to analyze the data. This is beyond the scope of the present work, which may be considered as a first step toward performing such a more complete model calculation.

The paper is organized as follows. In Sec.~\ref{Sec:formalism}, we introduce the framework of our theoretical model by explicitly giving the effective interaction Lagrangians, the resonance propagators, and the phenomenological form factors employed in this work. In Sec.~\ref{Sec:results}, our numerical results are shown and discussed. Finally, a brief summary and  conclusion of the present work is given in Sec.~\ref{sec:summary}.

\section{Formalism}  \label{Sec:formalism}

\begin{figure}[tbp]
\centering
{\vglue 0.15cm}
\subfigure[~$s$ channel]{
\includegraphics[width=0.45\columnwidth]{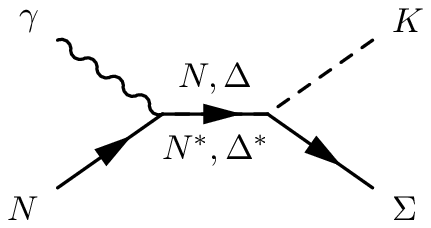}}  {\hglue 0.4cm}
\subfigure[~$t$ channel]{
\includegraphics[width=0.45\columnwidth]{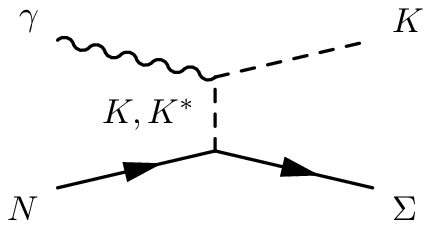}} \\[6pt]
\subfigure[~$u$ channel]{
\includegraphics[width=0.45\columnwidth]{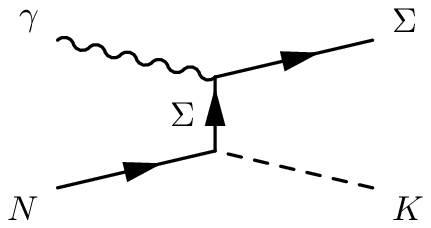}} {\hglue 0.4cm}
\subfigure[~Interaction current]{
\includegraphics[width=0.45\columnwidth]{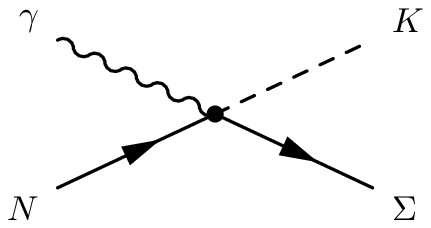}}
\caption{Generic structure of the amplitude for $\gamma N \to K \Sigma $. Time proceeds from left to right.}
\label{FIG:feymans}
\end{figure}

In the present work, we analyze the available data for $\gamma n \to K^+\Sigma^-$ within a tree-level effective Lagrangian approach. As diagrammatically depicted in Fig.~\ref{FIG:feymans}, the following diagrams are taken into account in constructing the reaction amplitude: (a) the $s$-channel $N$, $N^\ast$, $\Delta$, and $\Delta^\ast$ exchanges, (b) the $t$-channel $K$ and $K^\ast(892)$ exchanges, (c) the $u$-channel $\Sigma$ exchange, and (d) the interaction current. We mention that the $u$-channel $\Sigma^*$ exchange and $t$-channel $K_1$ exchange are not considered in the present work as we don't have experimental information on their radiative decays to constrain the corresponding coupling constants, and moreover, as can be seen in Fig.~\ref{fig:dsig}, the cross-section data for the considered reaction at high-energy backward angles and forward angles where the $\Sigma^*$ and $K_1$ exchanges contribute dominantly are scarce to fix the strengths of the exchanges of these two hadrons. The full photoproduction amplitude of $\gamma n \to K^+\Sigma^-$ can be written as \cite{Haberzettl:1997,Haberzettl:2006bn,Huang:2012,Huang:2012xj}
\begin{equation}
M^{\mu} \equiv M^{\mu}_s + M^{\mu}_t + M^{\mu}_u + M^{\mu}_{\rm int},  \label{eq:amplitude}
\end{equation}
with $\mu$ denoting the Lorentz index of the incoming photon $\gamma$. The first three terms in Eq.~(\ref{eq:amplitude}) represent the amplitudes arising from the $s$-, $t$-, and $u$-channel diagrams, respectively, and they can be straightforwardly calculated from the effective Lagrangians, propagators and form factors which will be given in detail in the following part of this section. The last term in Eq.~(\ref{eq:amplitude}), $M^{\mu}_{\rm int}$, denotes the interaction current which contains contributions from other diagrams that do not have $s$-, $t$-, or $u$-channel poles. An exact calculation of $M^{\mu}_{\rm int}$ is impractical as it includes very complicated non-linear interaction dynamics. In practice, we follow Refs.~\cite{Haberzettl:2006bn,Huang:2012xj} to model the interaction current $M^{\mu}_{\rm int}$ by a generalized contact current 
\begin{equation}
M^{\mu}_{\rm int} = \Gamma_{\Sigma NK}(q) C^\mu + M^{\mu}_{\rm KR} f_t, 
\label{eq:Mint}
\end{equation}
with $\Gamma_{\Sigma NK}(q)$ being the vertex function of $\Sigma NK$ interaction  obtained from the Lagrangian of Eq.~(\ref{eq:sigNK}),
\begin{equation}
\Gamma_{\Sigma NK}(q) = g_{\Sigma NK}\gamma_5 \left(\lambda + \frac{1-\lambda}{2M_N} {\slashed q}\right).
\end{equation}
Here $q$ is the four-momentum of the outgoing $K$ meson, and $\lambda$ is the mixing parameter for pseudoscalar ($\lambda=1$) and pseudovector ($\lambda=0$) $\Sigma NK$ couplings. $M^{\mu}_{\rm KR}$ in Eq.~(\ref{eq:Mint}) is the Kroll-Ruderman term obtained from the Lagrangian of Eq.~(\ref{eq:sigNKgamma}),
\begin{equation}
M^\mu_{KR} = -g_{\Sigma NK}\frac{1-\lambda}{2M_N} \gamma_5\gamma^\mu \tau Q_K,
\end{equation}
with $\tau$ denoting the isospin factor and $Q_K$ representing the electric charge of $K$ meson. $f_t$ in Eq.~(\ref{eq:Mint}) is the phenomenological form factor attaching to the $\Sigma NK$ vertex in $t$-channel $K$ exchange amplitude, and its form will be given in Eq.~(\ref{eq:ff_M}). $C^\mu$ in Eq.~(\ref{eq:Mint}) is an auxiliary current introduced to ensure that the full photoproduction amplitude given in Eq.~(\ref{eq:amplitude}) satisfies the generalized Ward-Takahashi identity and thus is fully gauge invariant. Following Refs.~\cite{Haberzettl:1997,Haberzettl:2006bn,Huang:2012,Huang:2012xj}, for the $\gamma n \to K^+\Sigma^-$ reaction, we choose the following prescription for $C^\mu$:
\begin{equation}
C^\mu =  - Q_{K} \tau \frac{f_t-\hat{F}}{t-q^2}  (2q-k)^\mu  - Q_{\Sigma} \tau \frac{f_{u}-\hat{F}}{u-p^{\prime 2}} (2p^{\prime}-k)^\mu,
\end{equation}
with
\begin{equation}
\hat{F} = 1 - \hat{h} \left(1 -  f_t\right) \left(1 -  f_u\right), 
\end{equation}
where $k$ and $p^\prime$ are the four-momenta for the incoming photon $\gamma$ and outgoing $\Sigma$, respectively; $Q_{\Sigma}$ is the electric charge of $\Sigma$; $f_u$ is the phenomenological form factor attaching to the $\Sigma NK$ vertex in $u$-channel $\Sigma$ exchange amplitude and will be given in Eq.~(\ref{eq:ff_B}); $\hat{h}=1$ is used for simplicity as usual \cite{Wang:2017tpe,Wang:2018vlv,Wei:2019}.

\subsection{Effective Lagrangians} \label{Sec:Lagrangians}

In this section, we present the explicit expressions of the Lagrangians employed in the present work. For the sake of brevity, we define the following operators 
\begin{equation}
\Gamma^{(+)}=\gamma_5 \qquad {\rm and} \qquad \Gamma^{(-)}=1,
\end{equation}
and the field-strength tensor of the photon field $A^\mu$
\begin{equation}
F^{\mu\nu} = \partial^\mu A^\nu - \partial^\nu A^\mu. 
\end{equation} 

The following Lagrangians are used for non-resonant production amplitudes: 
\begin{eqnarray}
{\cal L}_{\gamma KK} &=& ie \!\left[K^+\left(\partial_\mu K^-\right)-K^-\left(\partial_\mu K^+\right)\right] \! A^\mu,   \\[6pt]
{\cal L}_{\Sigma NK} &=& -\, g_{\Sigma NK}\bar{\Sigma}\gamma_{5} \left[\left(i\lambda + \frac{1-\lambda}{2M_N} \slashed{\partial}\right)K\right] N   \nonumber   \\
      && +\, \hc,  \label{eq:sigNK}  \\[6pt]
{\cal L}_{\gamma K{K^\ast}} &=& e\frac{g_{\gamma K{K^\ast}}}{M_K}\varepsilon^{\alpha \mu \lambda \nu}\left(\partial_\alpha A_\mu\right)\left(\partial_\lambda K\right) K^\ast_\nu,   \\[6pt]
{\cal L}_{\Sigma N {K^\ast}} &=&  -\, g_{\Sigma N {K^\ast}} \bar{\Sigma} \left[\left(\gamma^\mu-\frac{\kappa_{\Sigma N {K^\ast}}}{2M_N}\sigma^{\mu \nu}\partial_\nu\right)K^\ast_\mu\right] N \nonumber   \\
      && +\, \hc,  \\[6pt]
{\cal L}_{\Sigma \Sigma \gamma} &=& -\,e \bar{\Sigma}\left[\left(\hat{e}\gamma^{\mu} - \frac{\hat{\kappa}_{\Sigma}}{2M_N}\sigma^{\mu\nu}\partial_\nu \right)A_{\mu}\right]\Sigma,  \\[6pt] 
{\cal L}_{\gamma NN} &=& -\,e \bar{N} \!\left[ \! \left( \hat{e} \gamma^\mu - \frac{ \hat{\kappa}_N} {2M_N}\sigma^{\mu \nu}\partial_\nu \! \right) \! A_\mu\right]\! N,  \\[6pt]
%
%
%
{\cal L}_{\Delta\Sigma K} &=&  \frac{g_{\Delta\Sigma K}}{M_K} \bar{\Sigma} \left(\partial^{\mu}K\right) \Delta_{\mu} +\hc,       \\[6pt]
{\cal L}_{\Delta N\gamma} &=& -ie\frac{g^{(1)}_{\Delta N\gamma}}{2M_N}\bar{\Delta}_\mu \gamma_\nu \gamma_{5} F^{\mu\nu}N   \nonumber   \\
       && +\, e\frac{g^{(2)}_{\Delta N\gamma}}{(2M_N)^2}\bar{\Delta}_\mu  \gamma_{5} F^{\mu\nu} \partial_\nu N + \hc,   \\[6pt]  
L_{\Sigma NK \gamma} &=& ig_{\Sigma NK}\frac{1-\lambda}{2M_N}\bar{\Sigma}\gamma_5\gamma^\mu A_\mu \hat{Q}_K K\tau  N. \label{eq:sigNKgamma} 
\end{eqnarray}    
Here, $e$ stands for the elementary charge unit and $\hat{e}$ stands for the charge operator acting on the $N$ field or the $\Sigma$ field. $\hat{Q}_K$ is the charge operator acting on the $K$ field. $\hat{\kappa}_N \equiv \kappa_p\hat{e} + \kappa_n(1-\hat{e})$ with $\kappa_p=1.793$ being the anomalous magnetic moment of proton and $\kappa_n=-1.913$ the anomalous magnetic moment of neutron. $\hat{\kappa}_{\Sigma} \equiv \kappa_{\Sigma^+}(1+\hat{e})/2 + \kappa_{\Sigma^-}(1-\hat{e})/2$ with $\kappa_{\Sigma^+}=1.458$ being the anomalous magnetic moment of $\Sigma^+$ and $\kappa_{\Sigma^-}=-0.16$ the anomalous magnetic moment of $\Sigma^-$. $M_K$ and $M_N$ denote the masses of $K$ and $N$, respectively. $\varepsilon^{\alpha \mu \lambda \nu}$ is the totally antisymmetric Levi-Civita tensor with $\varepsilon^{0123}=+1$. In Eq.~(\ref{eq:sigNK}) $\lambda$ is the pseudoscalar-pseudovector mixing parameter for $\Sigma NK$ coupling and we use the pure pseudovector coupling, namely, $\lambda=0$ in our calculations. The coupling constant $g_{\gamma K^{\pm} K^{\ast\pm}} = 0.413$ is obtained by using the radiative decay width of $K^\ast\to K\gamma$ reported by RPP \cite{Zyla:2020zbs} with the sign determined by the flavor SU(3) symmetry \cite{Garcilazo:1993av}. The coupling constants $g^{(1)}_{\Delta N\gamma}=-4.18$ and $g^{(2)}_{\Delta N\gamma}=4.327$ are determined by the helicity amplitudes of $\Delta \to N \gamma$ from RPP \cite{Zyla:2020zbs}. The strong interaction coupling parameters are obtained via the flavor SU(3) symmetry \cite{Wang:2017tpe,Wang:2018vlv,Swart:1963,Ronchen:2013}:
\begin{eqnarray}
g_{\Sigma NK} &=& \frac{1}{5}g_{NN\pi}=2.692,  \\[6pt]  
g_{\Sigma NK^\ast} &=& \dfrac{1}{2} g_{NN\rho} -\dfrac{1}{2}g_{NN\omega}  = -4.26, \\[6pt]  
\kappa_{\Sigma N {K^\ast}} &=& \dfrac{f_{\Sigma NK^\ast}}{g_{\Sigma NK^\ast}}=\dfrac{f_{NN\rho}-f_{NN\omega}}{g_{NN\rho}-g_{NN\omega}}=-2.33,  \\[6pt] 
g_{\Delta\Sigma K} &=& \frac{M_K}{M_\pi} g_{\Delta N\pi} = 7.89,
\end{eqnarray} 
where the empirical values $g_{NN\pi}=13.46$, $g_{NN\rho}=3.25$, $g_{NN\omega}=11.76$, $\kappa_{NN\rho}=f_{NN\rho}/g_{NN\rho}=6.10$, $f_{NN\omega} = 0$, and $g_{\Delta N \pi}=2.23$ are used. We mention that the SU(3) relation is not ``exact" but a rough approximation. Its applicability needs to be tested by the capability of the model to describe the data.

For the $s$-channel $N^\ast$ and $\Delta^\ast$ exchanges, we use the following Lagrangians for the electromagnetic interactions
\begin{eqnarray}
{\cal L}_{RN\gamma}^{1/2\pm} &=& e\frac{g_{RN\gamma}^{(1)}}{2M_N}\bar{R} \Gamma^{(\mp)}\sigma_{\mu\nu} \left(\partial^\nu A^\mu \right) N  + \hc, \\[6pt]
{\cal L}_{RN\gamma}^{3/2\pm} &=& -\, ie\frac{g_{RN\gamma}^{(1)}}{2M_N}\bar{R}_\mu \gamma_\nu \Gamma^{(\pm)}F^{\mu\nu}N \nonumber \\
&&+\, e\frac{g_{RN\gamma}^{(2)}}{\left(2M_N\right)^2}\bar{R}_\mu \Gamma^{(\pm)}F^{\mu \nu}\partial_\nu N + \hc, \\[6pt]
{\cal L}_{RN\gamma}^{5/2\pm} & = & e\frac{g_{RN\gamma}^{(1)}}{\left(2M_N\right)^2}\bar{R}_{\mu \alpha}\gamma_\nu \Gamma^{(\mp)}\left(\partial^{\alpha} F^{\mu \nu}\right)N \nonumber \\
&& \pm\, ie\frac{g_{RN\gamma}^{(2)}}{\left(2M_N\right)^3}\bar{R}_{\mu \alpha} \Gamma^{(\mp)}\left(\partial^\alpha F^{\mu \nu}\right)\partial_\nu N \nonumber \\
&& + \,  \hc,  \\[6pt]
{\cal L}_{RN\gamma}^{7/2\pm} &=&  ie\frac{g_{RN\gamma}^{(1)}}{\left(2M_N\right)^3}\bar{R}_{\mu \alpha \beta}\gamma_\nu \Gamma^{(\pm)}\left(\partial^{\alpha}\partial^{\beta} F^{\mu \nu}\right)N \nonumber \\
&&-\, e\frac{g_{RN\gamma}^{(2)}}{\left(2M_N\right)^4}\bar{R}_{\mu \alpha \beta} \Gamma^{(\pm)} \left(\partial^\alpha \partial^\beta F^{\mu \nu}\right) \partial_\nu N  \nonumber \\
&&  + \,  \hc,
\end{eqnarray}
and use the following Lagrangians for the hadronic interactions
\begin{eqnarray} 
{\cal L}_{R\Sigma K}^{1/2\pm} &=& -g_{R\Sigma K} \bar{N}\Gamma^{(\pm)} \left[\left(i\lambda+\frac{1-\lambda}{M_R\pm M_\Sigma} {\slashed \partial}\right)K\right] R \nonumber   \\
      && +\, \hc, \label{eq:half1}    \\[6pt]
{\cal L}_{R\Sigma K}^{3/2\pm} &=& \pm \frac{g_{R\Sigma K}}{M_K} \bar{N}\Gamma^{(\mp)} \left(\partial^{\alpha}K\right) R_{\alpha} +\hc,      \\[6pt]
{\cal L}_{R\Sigma K}^{5/2\pm} &=& i \frac{g_{R\Sigma K}}{M_K^2} \bar{N}\Gamma^{(\pm)} \left(\partial^{\alpha}\partial^{\beta}K\right) R_{\alpha\beta} +\hc,      \\[6pt]
{\cal L}_{R\Sigma K}^{7/2\pm} &=& \mp\frac{g_{R\Sigma K}}{M_K^3} \bar{N}\Gamma^{(\mp)}\left(\partial^{\alpha}\partial^{\beta}\partial^{\gamma}K\right)R_{\alpha\beta\gamma}  \nonumber \\
      && + \, \hc, 
\end{eqnarray}
where $R$ denotes the $N^\ast$ or $\Delta^\ast$ resonance and the superscript of ${\cal L}$ denotes the spin and parity of the resonance $R$. The pseudoscalar-pseudovector mixing parameter $\lambda=0$ is employed in Eq.~(\ref{eq:half1}). When the resonance helicity amplitudes of $R\to N\gamma$ are available in RPP \cite{Zyla:2020zbs}, they will be used to fix the resonance electromagnetic couplings. Otherwise, the products of the hadronic coupling constant $g_{R\Sigma K}$ and the electromagnetic coupling constants $g_{RN\gamma}^{(i)}$ ($i=1,2$), which are relevant to the photoproduction amplitudes in the present work, will be determined by a fit to the data for $\gamma n \to K^+\Sigma^-$.

\subsection{Resonance propagators}

The following prescriptions for propagators of the resonances with spin $1/2$, $3/2$, $5/2$, and $7/2$ are employed in the present work \cite{Wang:2017tpe}:
\begin{eqnarray}
S_{1/2}(p) &=& \frac{i}{\slashed{p} - M_R + i \Gamma_R/2}, \label{propagator-1hf}  \\[6pt]
S_{3/2}(p) &=&  \frac{i}{\slashed{p} - M_R + i \Gamma_R/2} \left( \tilde{g}_{\mu \nu} + \frac{1}{3} \tilde{\gamma}_\mu \tilde{\gamma}_\nu \right),  \\[6pt]
S_{5/2}(p) &=&  \frac{i}{\slashed{p} - M_R + i \Gamma_R/2} \,\bigg[ \, \frac{1}{2} \big(\tilde{g}_{\mu \alpha} \tilde{g}_{\nu \beta} + \tilde{g}_{\mu \beta} \tilde{g}_{\nu \alpha} \big)  \nonumber \\
&& -\, \frac{1}{5}\tilde{g}_{\mu \nu}\tilde{g}_{\alpha \beta}  + \frac{1}{10} \big(\tilde{g}_{\mu \alpha}\tilde{\gamma}_{\nu} \tilde{\gamma}_{\beta} + \tilde{g}_{\mu \beta}\tilde{\gamma}_{\nu} \tilde{\gamma}_{\alpha}  \nonumber \\
&& +\, \tilde{g}_{\nu \alpha}\tilde{\gamma}_{\mu} \tilde{\gamma}_{\beta} +\tilde{g}_{\nu \beta}\tilde{\gamma}_{\mu} \tilde{\gamma}_{\alpha} \big) \bigg], \\[6pt]
S_{7/2}(p) &=&  \frac{i}{\slashed{p} - M_R + i \Gamma_R/2} \, \frac{1}{36}\sum_{P_{\mu} P_{\nu}} \bigg( \tilde{g}_{\mu_1 \nu_1}\tilde{g}_{\mu_2 \nu_2}\tilde{g}_{\mu_3 \nu_3} \nonumber \\
&& -\, \frac{3}{7}\tilde{g}_{\mu_1 \mu_2}\tilde{g}_{\nu_1 \nu_2}\tilde{g}_{\mu_3 \nu_3} + \frac{3}{7}\tilde{\gamma}_{\mu_1} \tilde{\gamma}_{\nu_1} \tilde{g}_{\mu_2 \nu_2}\tilde{g}_{\mu_3 \nu_3} \nonumber \\
&& -\, \frac{3}{35}\tilde{\gamma}_{\mu_1} \tilde{\gamma}_{\nu_1} \tilde{g}_{\mu_2 \mu_3}\tilde{g}_{\nu_2 \nu_3} \bigg),  \label{propagator-7hf}
\end{eqnarray}
with $\tilde{g}_{\mu \nu}$ and $\tilde{\gamma}_{\mu}$ being defined as
\begin{eqnarray}
\tilde{g}_{\mu \nu} &=& -\, g_{\mu \nu} + \frac{p_{\mu} p_{\nu}}{M_R^2}, \\[6pt]
\tilde{\gamma}_{\mu} &=& -\, \gamma_{\mu} + \frac{p_{\mu}\slashed{p}}{M_R^2}.   \label{eq:prop-auxi}
\end{eqnarray}
Here $M_R$ and $\Gamma_R$ are the mass and width for the resonance $R$ with four-momentum $p$, respectively. The summation over $P_\mu$ $\left(P_\nu\right)$ in Eq.~(\ref{propagator-7hf}) goes over all the $3!=6$ possible permutations of the indices $\mu_1\mu_2\mu_3$ $\left(\nu_1\nu_2\nu_3\right)$.

\subsection{Form factors} \label{Sec:Regge treatment}

In the present work, each hadronic vertex obtained from the Lagrangians given in Sec.~\ref{Sec:Lagrangians} is accompanied with a phenomenological form factor. For $t$-channel meson exchanges, the following form factor is employed \cite{Huang:2012xj,Wang:2017tpe,Wang:2018vlv}: 
\begin{eqnarray}
f_M(q^2_M) =  \left (\frac{\Lambda_M^2-M_M^2}{\Lambda_M^2-q^2_M} \right)^2,   \label{eq:ff_M}
\end{eqnarray}
where $M_M$ and $q_M$ denote the mass and four-momentum of the intermediate meson, respectively, and $\Lambda_M$ is the cut off parameter for meson exchange diagrams. For $s$-channel and $u$-channel baryon exchanges, we use the following form factor \cite{Huang:2012xj,Wang:2017tpe,Wang:2018vlv}:
\begin{eqnarray}
f_B(p^2_x) =\left (\frac{\Lambda_B^4}{\Lambda_B^4+\left(p_x^2-M_B^2\right)^2} \right )^2,  \label{eq:ff_B}
\end{eqnarray}
where $M_B$ and $p_x$ ($x=s, u$) represent the mass and four-momentum of the exchanged baryon in $s$ or $u$ channel, respectively, and $\Lambda_B$ is the cut off parameter for baryon exchange diagrams.

\section{Results and discussion}   \label{Sec:results}

\begin{table*}[htb]
\caption{Fitted values of resonant parameters and the extracted resonance decay branching ratios. The asterisks below the resonance names denote the overall rating of each resonance evaluated by RPP \cite{Zyla:2020zbs}. The values in the brackets blow resonances' masses, widths, and branching ratios are corresponding values advocated by RPP \cite{Zyla:2020zbs}. }
\begin{tabular*}{\textwidth}{@{\extracolsep\fill}ccccccc}
\hline\hline
 & $M_R$ [MeV] &$\Gamma_R$ [MeV] &$g^{(1)}_{RN\gamma} g_{R\Sigma K}$ &$g^{(2)}_{RN\gamma} g_{R\Sigma K}$ &$\beta_{N\gamma}$ [\%] &$\beta_{K\Sigma}$ [\%]  \\
\hline
$N(1710)1/2^+$   & $1692 \pm 1$  &$161 \pm 1$    &$1.661 \pm 0.012$    &&$\bm{0.01}$ &$0.008$      \\
$\ast\!\ast\!\ast\ast$  & $[1680\sim 1740]$   &$[80\sim 200]$ &&&$[0.0\sim 0.02]$   & [seen]     \\   \hline
$N(1880)1/2^+$  & $1893 \pm 1$  &$400 \pm 2$    &$1.252 \pm 0.001$    &&$\bm{0.05}$ &$18.531$       \\
$\ast\!\ast\!\ast$      & $[1830\sim 1930]$   &$[200\sim 400]$&&&$[0.002\sim 0.63]$ &$[10\sim 24]$    \\   \hline
$N(1895)1/2^-$  & $1870 \pm 1$  &$200 \pm 1$    &$0.054 \pm 0.001$    &&$\bm{0.027}$ &$8.795$       \\
$\ast\!\ast\!\ast\ast$  & $[1870\sim 1920]$   &$[80\sim 200]$ &&&$[0.003\sim 0.05]$ &$[6\sim 20]$     \\   \hline      
$N(1900)3/2^+$  & $1942 \pm 1$  &$122 \pm 2$ &$-0.116 \pm 0.001$ &$0.024 \pm 0.004$ &$\bm{0.04}$ &$6.999$       \\
$\ast\!\ast\!\ast\ast$  & $[1890\sim 1950]$ &$[100\sim 320]$  &&&$[\textless 0.04]$ &$[3\sim 7]$     \\   \hline      
$N(2060)5/2^-$  & $2030 \pm 2$  &$450 \pm 2$ &$-0.648 \pm 0.023$  &$0.254 \pm 0.034$ &$\bm{0.036}$ &$1.024$       \\
$\ast\!\ast\!\ast$      & $[2030\sim 2200]$ &$[300\sim 450]$  &&&$[0.003 \sim 0.07]$ &$[1\sim 5]$     \\   \hline   
$\Delta(1910)1/2^+$    & $1913 \pm 2$  &$200 \pm 1$  &$0.541 \pm 0.006$  && $\bm{0.01}$ &$13.980$       \\
$\ast\!\ast\!\ast\ast$  & $[1850\sim 1950]$ &$[200\sim 400]$  &&& $[0.0 \sim 0.02]$ &$[4\sim 14]$      \\   \hline   
$\Delta(1920)3/2^+$ & $1870 \pm 1$ &$360 \pm 1$ &$-1.204 \pm 0.003$ &$-0.103 \pm 0.034$ &$\bm{0.2}$ &$3.019$    \\
$\ast\!\ast\!\ast$      & $[1870\sim 1970]$ &$[240\sim 360]$  &&&   &$[2\sim 6]$     \\   \hline                                                                    
$\Delta(1950)7/2^+$ & $1915 \pm 1$ &$235 \pm 5$ &$7.416 \pm 0.220$ &$-7.261 \pm 0.215$ & $\bm{0.001}$ &$0.300$          \\
$\ast\!\ast\!\ast\ast$  & $[1915\sim 1950]$ &$[235\sim 335]$  &   && &$[0.3\sim 0.5]$     \\   \hline \hline                   
\end{tabular*}
\label{tab:resonant parameters}
\end{table*}

\begin{table*}[htb]
\caption{Fitted values of cutoff parameters (in MeV).}
\begin{tabular*}{\textwidth}{@{\extracolsep\fill}ccccccccccc}
\hline \hline
$\Lambda_{N(1710)}$ & $\Lambda_{N(1880)}$ & $\Lambda_{N(1895)}$ & $\Lambda_{N(1900)}$ & $\Lambda_{N(2060)}$ & $\Lambda_{\Delta(1910)}$ & $\Lambda_{\Delta(1920)}$ & $\Lambda_{\Delta(1950)}$ & $\Lambda_{K^\ast(K)}$  &  $\Lambda_{\Delta(N)}$  & $\Lambda_{\Sigma}$  \\ \hline
$1932 \pm 9$ & $980 \pm 5$ &  $982 \pm 16$ & $2000 \pm 8$ & $1620 \pm 8$ & $1222 \pm 25$ & $1317 \pm 6$ & $800 \pm 1$ &   $584 \pm 1$             &  $1140 \pm 2$               &  $1092 \pm 8$  \\ \hline
\hline
\end{tabular*}
\label{tab:cutoff_parameters}
\end{table*}

\begin{figure*}[htb]
\includegraphics[width=1.0\textwidth]{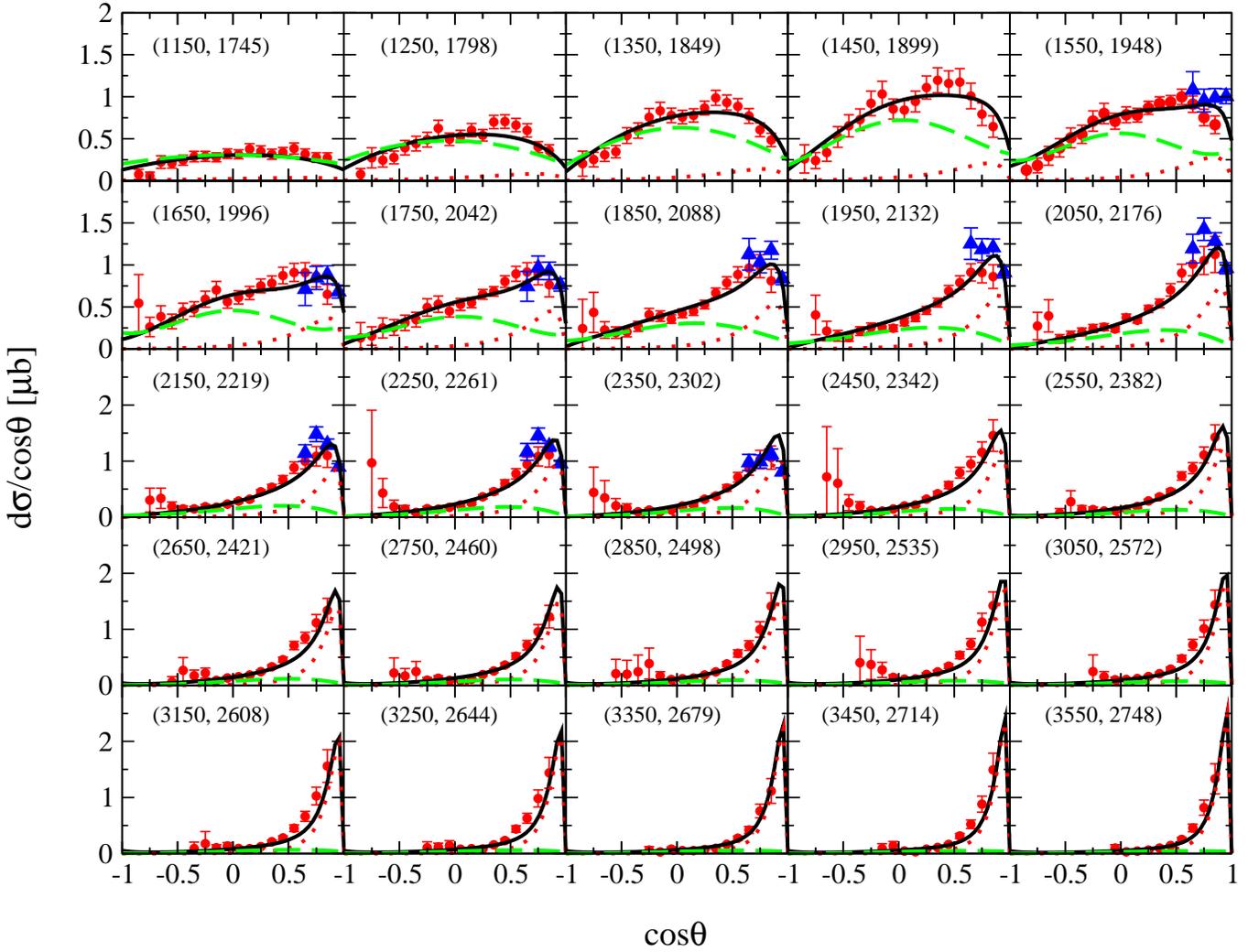}
\caption{Differential cross sections for $\gamma n \to K^+\Sigma^-$. The solid (black), dotted (red), and long-dashed (green) lines represent the results from the full calculation, the $t$-channel amplitudes ($K$ and $K^\ast(892)$ exchanges), and $s$-channel amplitudes ($N$, $N^\ast$, $\Delta$, and $\Delta^\ast$ exchanges), respectively. Data with red circles and blue triangles are taken from the CLAS Collaboration \cite{AnefalosPereira:2009zw} and the LEPS Collaboration \cite{Kohri:2006yx}, respectively. The numbers in parentheses denote the incident energies (left number) and the corresponding center-of-mass energies of the system (right number), in MeV. }
\label{fig:dsig}
\end{figure*}
 
\begin{figure*}[htb]
\includegraphics[width=1.0\textwidth]{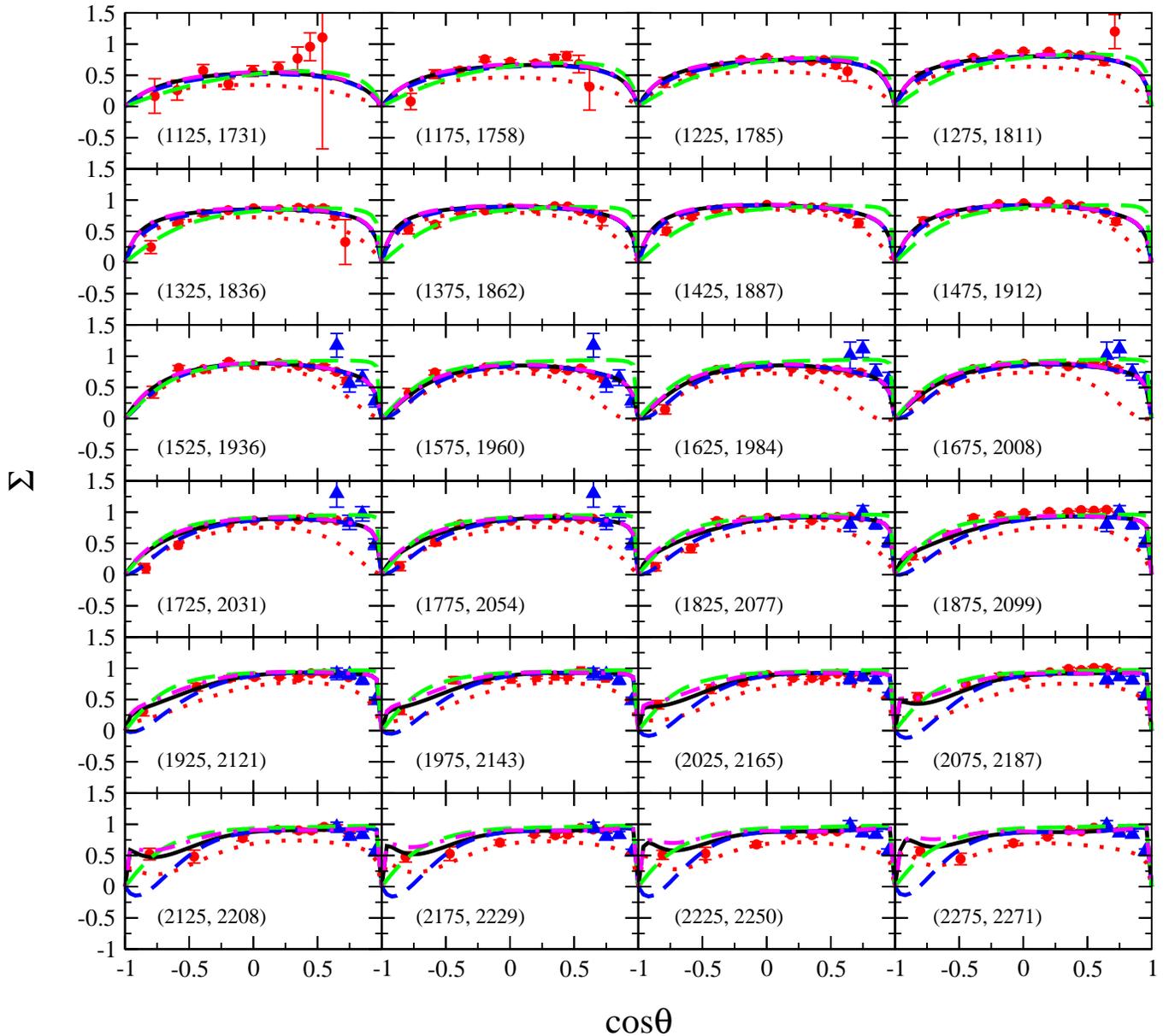}
\caption{Photo-beam asymmetries $\Sigma$ for $\gamma n \to K^+\Sigma^-$. The solid (black) lines represent the results from the full calculation. The dotted (red), dashed (blue), long-dashed (green), and dot-dashed (magenta) lines represent the results obtained by switching off the $t$-channel amplitudes ($K$ and $K^\ast(892)$ exchanges), $u$-channel amplitude ($\Sigma$ exchange), $s$-channel amplitudes ($N$, $N^\ast$, $\Delta$, and $\Delta^\ast$ exchanges), and interaction current, respectively. Data with red circles and blue triangles are taken from the CLAS Collaboration \cite{CLAS-beam} and the LEPS Collaboration \cite{Kohri:2006yx}, respectively. The numbers in parentheses denote the incident energies (left number) and the corresponding center-of-mass energies of the system (right number), in MeV.}
\label{fig:beam}
\end{figure*}

\begin{figure}[htb]
\includegraphics[width=\columnwidth]{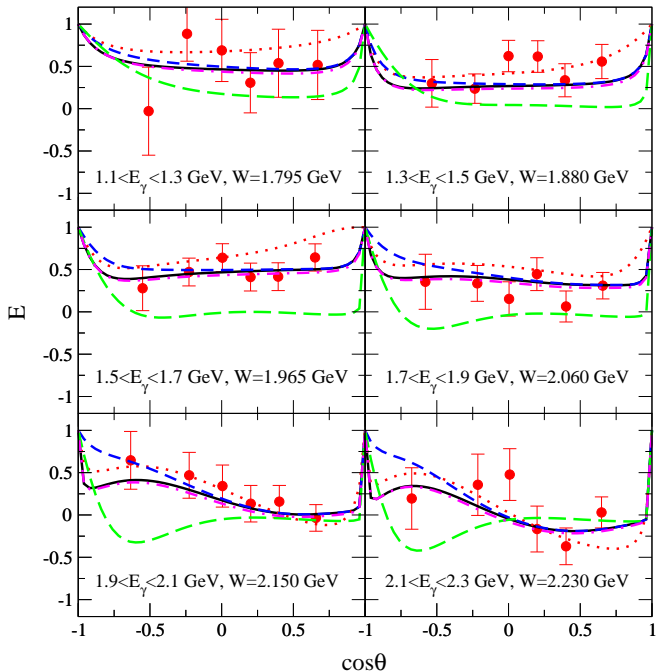}
\caption{Beam-target asymmetries $E$ for $\gamma n \to K^+\Sigma^-$. Notations for lines are the same as Fig.~\ref{fig:beam}. Data with red circles are taken from the CLAS Collaboration \cite{Zachariou:2020kkb}. $E_\gamma$ and $W$ denote the incident energies and the corresponding center-of-mass energies of the system, respectively.}
\label{fig:e}
\end{figure}

\begin{figure*}[htb]
\centering
\includegraphics[width=\columnwidth]{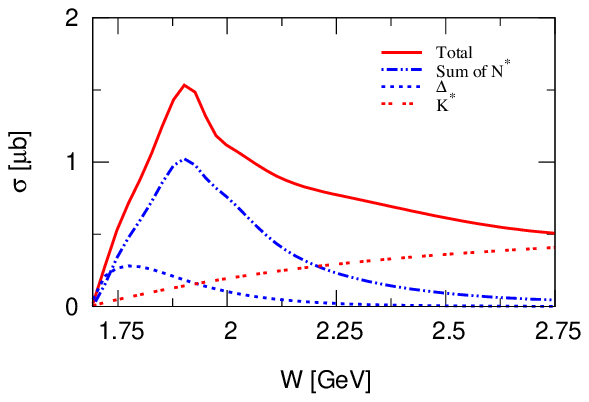}
\hspace{0.1 in}
\includegraphics[width=\columnwidth]{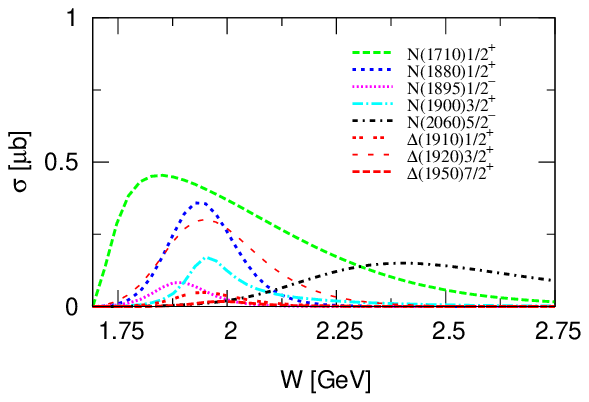}
\caption{Predicted total cross sections for $\gamma n \to K^+\Sigma^-$ with contributions from individual terms.}
\label{fig:total_cs}
\end{figure*}

The main purpose of the present work is to perform a combined analysis of all the available data, especially the very recently released photo-beam asymmetry data $\Sigma$ and the beam-target asymmetry data $E$,  for the $\gamma n \to K^+\Sigma^-$ reaction in an effective Lagrangian approach as introduced in Sec.~\ref{Sec:formalism}. We consider the $s$-channel $N$, $N^\ast$, $\Delta$, and $\Delta^\ast$ exchanges, the $t$-channel $K$ and $K^\ast(892)$ exchanges, the $u$-channel $\Sigma$ exchange, and the interaction current in constructing the reaction amplitudes.

As can be seen from Fig.~\ref{fig:dsig}, the differential cross-section data for $\gamma n \to K^+\Sigma^-$ from the CLAS Collaboration \cite{AnefalosPereira:2009zw} indicate possible contributions from the $s$-channel resonance exchanges in the energy range from the $\Sigma K$ threshold up to $E_\gamma \sim 1850$ MeV ($W \sim 2088$ MeV). In the most recent RPP \cite{Zyla:2020zbs}, there are in total seven $N$ and $\Delta$ resonances with masses $M \leq 2100$ MeV that have considerable $K \Sigma$ branching ratios, namely, the $N(1880)1/2^+$, $N(1895)1/2^-$, $N(1900)3/2^+$, $N(2060)5/2^-$, $\Delta(1910)1/2^+$, $\Delta(1920)3/2^+$, and $\Delta(1950)7/2^+$ resonances. Among these seven resonances, the $N(1895)1/2^-$, $N(1900)3/2^+$, $\Delta(1910)1/2^+$, and $\Delta(1950)7/2^+$ are rated as four-star resonances, and the other three as three-star resonances in RPP \cite{Zyla:2020zbs}. As a first step, we try to see if the available experimental data for $K^+\Sigma^-$ photoproduction can be well described by considering in the $s$ channel the contributions from these seven resonances. Their masses and widths will be varied within the ranges advocated by RPP \cite{Zyla:2020zbs} to fit the data for $\gamma n \to K^+\Sigma^-$.

We use MINUIT to fit the model parameters, and numerous different starting values of the model parameters have been tested to achieve the best converged fitted results. It turns out, however, that with only the seven resonances mentioned above being taken into account in the calculation, we were unable to obtain an acceptable description of the available data for $\gamma n \to K^+\Sigma^-$. In particular, the near-threshold angular distribution data cannot be well reproduced. This hints for contribution of additional resonance(s) located near this reaction threshold. 

Actually, in addition to the above-mentioned seven resonances which have considerable $K \Sigma$ branching ratios, in the $K\Sigma$ near threshold region, there is another four-star resonance, i.e. the $N(1710)1/2^+$, that is marked as ``seen" in its decay branching ratio to the $K \Sigma$ channel in RPP \cite{Zyla:2020zbs}. We then try to refit the model parameters by further including this resonance in our model. It turns out that a satisfactory description of all the available data for $\gamma n \to K^+\Sigma^-$ can be obtained, as we shall show and discuss in detail in the following. 

As a result, we include five nucleon resonances and three $\Delta$ resonances with 38 fitting parameters for resonances and background contributions in our model. These parameters are determined by a fit to 780 available data for $K^+ \Sigma^-$ photoproduction within the energy range from threshold up to $E_\gamma = 3550$ MeV. The resulted $\chi^2$ per degree of freedom is $\chi^2/N = 3.11$. In Table~\ref{tab:resonant parameters} we list the fitted values of resonance parameters and the extracted resonance branching ratios. There, the number of asterisks below the resonance names denote the overall status of each resonance evaluated by RPP \cite{Zyla:2020zbs}. The values in the brackets below the resonance masses and widths denote the ranges of the corresponding values given by RPP \cite{Zyla:2020zbs}. As for the resonance couplings, since the reaction amplitudes are only sensitive to the products of the electromagnetic and hadronic coupling constants, we show in Table~\ref{tab:resonant parameters} these products instead of individual electromagnetic and hadronic coupling constants. For the four-star resonances $\Delta(1910)1/2^+$ and $\Delta(1950)7/2^+$, their helicity amplitudes are available in RPP, which can be used to fix the resonance electromagnetic coupling constants. The results give $g_{RN\gamma}^{(1)} = -0.103$ for the $\Delta(1910)1/2^+$ resonance, and $g_{RN\gamma}^{(1)} = -12.92$, $g_{RN\gamma}^{(2)} = 12.65$ for the $\Delta(1950)7/2^+$ resonance. Hence, the hadronic coupling constants $g_{R\Sigma K} = -5.252$ for $\Delta(1910)1/2^+$ and $g_{R\Sigma K} = -0.574$ for $\Delta(1950)7/2^+$ can be extracted from the products of hadronic and electromagnetic couplings listed in Table~\ref{tab:resonant parameters}. For other resonances, the helicity amplitudes are not available in RPP. The last two columns in Table~\ref{tab:resonant parameters} show the extracted decay branching rations of $R \to N\gamma$ and $R \to K\Sigma$, respectively. The values in  brackets denote the ranges of the corresponding branching ratios available in RPP \cite{Zyla:2020zbs}. The electromagnetic branching ratios in bold fonts are fixed and the hadronic branching ratios are then extracted from the products of the resonance electromagnetic and hadronic coupling constants. For the $\Delta(1910)1/2^+$ and $\Delta(1950)7/2^+$ resonances, the electromagnetic partial decay widths are calculated from their electromagnetic coupling constants, and for the $N(1710)1/2^+$, $N(1895)1/2^-$, and $N(2060)5/2^-$ resonances, the electromagnetic branching ratios are fixed to be the centroid values of the corresponding ranges given by RPP. For the $N(1880)1/2^+$ resonance, a value $\beta_{N\gamma}=0.05\%$ which is smaller than the centroid RPP value of $0.316\%$ but still within the range quote by RRP is used to avoid that the corresponding hadronic branching ratio to fall well outside its range given in RPP. For the same reason, a value $\beta_{N\gamma}=0.04\%$ is adopted for the $N(1900)3/2^+$ resonance. For the $\Delta(1920)3/2^+$ resonance, there is no information about its electromagnetic branching ratio from RPP and a value $\beta_{N\gamma}=0.2\%$ is adopted to extract the hadronic branching ratio. One can see from Table~\ref{tab:resonant parameters} that, for all the resonances, the electromagnetic branching ratios and hadronic branching ratios are all consistent to those advocated by RPP \cite{Zyla:2020zbs}. The fitted values of cut-off parameters for resonances and background diagrams are given in Table~\ref{tab:cutoff_parameters}. Note that, based on the facts that the contributions of $K$ and $N$ exchanges are rather small and the results are insensitive to the values of $\Lambda_K$ and $\Lambda_N$ when both of these two cutoffs are treated as free parameters, we simply set $\Lambda_K=\Lambda_{K^*}$ and $\Lambda_N=\Lambda_\Delta$ to reduce the number of fitting parameters. The values of the cut-off parameters for resonances, which are determined by fitting the data, vary in a wide range. As will be shown in Fig.~\ref{fig:total_cs}, a small cut-off means the s-channel resonance contributes in a narrow energy range, while a big value of the cut-off allows contributions at large energies.

We remark that a proper extraction of the resonances parameters should be done, in principle, by searching for the poles of the reaction amplitude in the complex energy plane \cite{Doring2009PLB,Doring2009NPA}. In particular, the resonance coupling constants are related to the residues at those poles. This is beyond the scope of the present work which is a first step toward such a more complete analysis.

In Fig.~\ref{fig:dsig}, our numerical results for differential cross sections (black-solid lines) are shown and compared to the data from the CLAS \cite{AnefalosPereira:2009zw} Collaboration (red circles) and LEPS \cite{Kohri:2006yx} Collaboration (blue triangle). Here the numbers in parentheses denote the incident energies (left number) and the corresponding center-of-mass energies of the system (right number), in MeV. Note that the CLAS data are measured for $K^+\Sigma^-$ photoproduction off quasi-free neutron. We have checked and found that the Fermi motion effect here is very tiny and thus is ignored in our calculations. One sees that, overall, the model results agree with the data quite well in the whole energy region considered. The contributions from individual terms are also shown. The red-dotted and green-long-dashed lines represent the results from the $t$-channel amplitudes ($K$ and $K^\ast(892)$ exchanges) and $s$-channel amplitudes ($N$, $N^\ast$, $\Delta$, and $\Delta^\ast$ exchanges), respectively. It is seen that in the lower energy region, the contributions from the $s$-channel mechanisms dominate the differential cross sections of the $K^+\Sigma^-$ photoproduction. As the energy increases, the contributions from the $t$-channel mechanism become more and more prominent and are essential in describing the data at forward angles. Furthermore, the $K^\ast$ exchange is largely responsible for the forward-angle-peaked behavior exhibited at higher energies. This explains that $\Lambda_{K^\ast(K)}$, which is the only parameter in the $t$-channel amplitudes, is well determined to be $\Lambda_{K^\ast(K)} = 584 \pm 1$ MeV as shown in Table~\ref{tab:cutoff_parameters}. Note that the data seem to show a backward-angle enhancement in the energy region of $E_\gamma=2250-2450$ MeV, although they have large error bars. The model results seem to underestimate these data, but they are still within the experimental error range. The contributions of $u$-channel mechanism and the interaction current are negligible to the differential cross sections and they are not plotted out in this figure. We mention that, in principle, a stronger contribution at backward angles could be obtained if one includes the exchange of one more $\Sigma^\ast$ resonance in the $u$-channel. But in this case, one needs to introduce extra parameters as both the theoretical and experimental information on $\Sigma^\ast N \gamma$ interaction are quite scarce. These parameters cannot be well determined at present due to the large error bars of the backward-angle differential cross-section data. We postpone the inclusion of more $u$-channel resonances till more precision data for this reaction become accessible.

Our numerical results for beam asymmetries $\Sigma$ and beam-target asymmetries $E$ are shown and compared to the experimental data \cite{Kohri:2006yx,CLAS-beam,Zachariou:2020kkb} in Figs.~\ref{fig:beam} and \ref{fig:e}, respectively. Here, the black-solid lines represent the results of the full reaction amplitude. One can see that the overall agreement of the theoretical predictions with the data is rather reasonable. The red-dotted, blue-dashed, green-long-dashed, and magenta-dot-dashed lines represent the results obtained by switching off the $t$-channel amplitudes ($K$ and $K^\ast(892)$ exchanges), $u$-channel amplitude ($\Sigma$ exchange), $s$-channel amplitudes ($N$, $N^\ast$, $\Delta$, and $\Delta^\ast$ exchanges), and the interaction current, respectively. For  photo-beam asymmetries $\Sigma$ in Fig.~\ref{fig:beam}, one sees that the $s$-and the $t$-channel contributions (green-long-dashed lines and red-dotted lines, respectively) have significant effects on the photo-beam asymmetries. Noticeable influence of the $u$-channel contribution (blue-dashed lines) can be also seen at higher energies and backward angles. The contribution from the interaction current is negligible, similar to the situation encountered in the angular distributions as shown in Fig.~\ref{fig:dsig}. For the beam-target asymmetries $E$ in Fig.~\ref{fig:e}, one sees that the dominant contribution is coming from the $s$-channel resonance exchanges. The $t$- and $u$-channels also contribute significantly. The contribution from the interaction current is negligible, same as the situation seen in the differential cross sections and photo-beam asymmetries $\Sigma$ as shown in Figs.~\ref{fig:dsig} and \ref{fig:beam}, respectively.

The prediction of the total cross section for $\gamma n \to K^+\Sigma^-$ is shown in Fig.~\ref{fig:total_cs} (left panel), where its dynamical content is also exhibited. It is clearly seen that the contributions from the $s$-channel resonance exchanges dominate the total cross sections in the lower energy region and are responsible for the bump structure exhibited around $W\sim 1900$ MeV. As displayed in Fig.~\ref{fig:total_cs} (right panel), the resonance contributions are dominated by the $N(1710)1/2^+$ resonance, followed by the $N(1880)1/2^+$, $\Delta(1920)3/2^+$, and $N(1900)3/2^+$ resonances. The $N(2060)5/2^-$ resonance also contributes considerably at energies above $W \sim 2100$ MeV, while the contributions from other resonances are rather small. Note that the $N(1710)1/2^+$ resonance contributes in a much wider energy range with a stronger strength than the $N(1880)1/2^+$ resonance. This is because that the $N(1710)1/2^+$ has a much bigger cutoff and a bigger coupling constant than the $N(1880)1/2^+$, as listed in Tables~\ref{tab:resonant parameters} and \ref{tab:cutoff_parameters}. Nevertheless, the $N(1710)1/2^+$ resonance has a much smaller branching ratio into $K\Sigma$ than the $N(1880)1/2^+$ resonance. The reason is that the fitted mass of $N(1710)1/2^+$ is rather close to the $K\Sigma$ threshold, resulting in a much small phase space which suppresses the resonance decay width to $K\Sigma$ channel. Getting back to Fig.~\ref{fig:total_cs} (left panel), the $s$-channel $\Delta$ exchange provides a significant contribution in the low energy region, while the $t$-channel $K^\ast$ exchange provides significant contribution at higher energies. The contributions from other non-resonant diagrams to the total cross sections are negligible, and they are not plotted in Fig.~\ref{fig:total_cs}.

In Ref.~\cite{Byd2021}, the data for $\gamma n \to K^+ \Sigma^-$ were analyzed within an isobar model. There are two sets of fit results, i.e. fit M and fit L, which were obtained by using, respectively, the standard MINUIT and the so-called LASSO method for fitting the data. The latter method, together with an information criterium, allows a blindfold determination of the relevant baryon resonances by the data from a given set of resonances considered within a specified model. Compared with our present work, in fit M of Ref.~\cite{Byd2021}, an additional $K_1(1270)$ meson exchange was considered in the $t$ channel, and six more nucleon resonances in the $s$ channel, namely, the $N(1535)1/2^-$, $N(1650)1/2^-$, $N(1675)5/2^-$, $N(1720)3/2^+$, $N(1875)3/2^-$, and $N(2120)3/2^-$ resonances. For $\Delta$ resonances, the $\Delta(1900)1/2^-$ exchange was considered, while the $\Delta(1910)1/2^+$, $\Delta(1920)3/2^+$, and $\Delta(1950)7/2^+$ resonances which are included in our present work were not considered in Ref.~\cite{Byd2021}.
In fit L of Ref.~\cite{Byd2021}, the $K_1(1270)$ meson exchange in the $t$ channel was not considered.  Apart from the three nucleon resonances, $N(1895)1/2^-$, $N(1900)3/2^+$ and $N(2060)5/2^-$, also found in the present work, the fit L also finds the $N(1650)1/2^-$, $N(1675)5/2^-$, $N(1720)3/2^+$, $N(1875)3/2^-$, and $N(2120)3/2^-$ resonances to be relevant. In contrast, in the present work we include the $N(1710)1/2^+$, $N(1880)1/2^+$, $\Delta(1910)1/2^+$, $\Delta(1920)3/2^+$, and $\Delta(1950)7/2^+$ resonances, in addition to those three nucleon resonances mentioned above which are also found in fit L of Ref.~\cite{Byd2021}. Note that in our present work, the resonances are selected in such a way that only the resonances that have considerable branching ratios to $K\Sigma$ channel in RPP \cite{Zyla:2020zbs} are considered, while in Ref.~\cite{Byd2021} the resonances were determined by a fit to the data for the $\gamma n \to K^+\Sigma^-$ reaction. The differential cross sections and the photo-beam asymmetries $\Sigma$ for $\gamma n \to K^+ \Sigma^-$ were well described in Ref.~\cite{Byd2021}, but their description of the data on beam-target asymmetries $E$ were worse than the description in the present work. We mention that the data on beam-target asymmetries $E$ have not been included in the fits of Ref.~\cite{Byd2021}. It was concluded in Ref.~\cite{Byd2021} that the $s$-channel $N(1720)3/2^+$ resonance exchange is very important to describe the data. This conclusion was reached based on the fact that the $N(1720)3/2^+$ resonance exchange had big influence in reproducing the photo-beam asymmetry data (see Fig.~7 and Fig.~8 of Ref.~\cite{Byd2021}).   

The fact that the two independent analyses of Ref.~\cite{Byd2021} and the present work have led to different resonance contents in describing the same data and have drawn different conclusions about the reaction mechanisms, indicates that the currently available data  for $\gamma n \to K^+ \Sigma^-$ are insufficient to uniquely determine the reaction amplitude in theoretical models. More data are needed to further learn about the resonance content in this reaction. In fact, we mention that a pseudoscalar-meson photoproduction process requires at least eight carefully chosen independent observables to determine the corresponding reaction amplitude up to an overall phase \cite{Chiang:1996em,Nakayama:2019cys,Nakayama:2018yzw}.

\section{Summary and conclusion}  \label{sec:summary}

Quite recently, the CLAS Collaboration has released the data on beam asymmetries $\Sigma$ and beam-target asymmetries $E$ for the $\gamma n \to K^+\Sigma^-$ reaction \cite{CLAS-beam,Zachariou:2020kkb}. Stimulated by these new measurements, in the present work, we have performed an analysis of all the available data, including the differential cross sections \cite{AnefalosPereira:2009zw,Kohri:2006yx}, the beam asymmetries $\Sigma$ \cite{CLAS-beam,Kohri:2006yx}, and the beam-target asymmetries $E$ \cite{Zachariou:2020kkb}, for the reaction $\gamma n \to K^+\Sigma^-$ in an effective Lagrangian approach. The purpose is to understand the reaction mechanisms and to extract the information on the $N$ and $\Delta$ resonances involved. 

We have constructed the reaction amplitude for $\gamma n \to K^+\Sigma^-$ by considering the $t$-channel $K$ and $K^\ast(892)$ exchanges, the $u$-channel $\Sigma$ exchange, the $s$-channel $N$, $N^\ast$, $\Delta$, and $\Delta^\ast$ exchanges, and the interaction current. The gauge invariance of the production amplitude, dictated by the generalized Ward-Takahashi identity, has been strictly implemented. The $N(1710)1/2^+$, $N(1880)1/2^+$, $N(1895)1/2^-$, $N(1900)3/2^+$, $N(2060)5/2^-$, $\Delta(1910)1/2^+$, $\Delta(1920)3/2^+$, and $\Delta(1950)7/2^+$ resonances are considered in the $s$ channel with their masses and widths been varied within the ranges advocated by the most recent RPP \cite{Zyla:2020zbs} to fit the data. 

All the available data for $\gamma n \to K^+\Sigma^-$ are well reproduced. The present analysis reveals that the contributions from the $s$-channel mechanisms dominate the $\gamma n \to K^+\Sigma^-$ reaction in the lower energy region. The most prominent resonance contributions come from the $N(1710)1/2^+$, $N(1880)1/2^+$, $N(1900)3/2^+$, and $\Delta(1920)3/2^+$ resonances. The $N(2060)5/2^-$ resonance contributes considerably at energies above $W \sim 2100$ MeV. The contributions from the $t$-channel $K^\ast$ exchange are crucial to reproduce the differential cross-section data at forward angles in the higher energy region and are also significant to the photo-beam asymmetries $\Sigma$ and beam-target asymmetries $E$. The interaction current turns out to contribute slightly for all the observables considered, namely, the cross sections, beam asymmetries $\Sigma$, and beam-target asymmetries $E$ of the $\gamma n \to K^+\Sigma^-$ reaction. However, the resonance content in this reaction is far from being understood. In this regard, more independent data on this reaction are required.

A combined analysis of all the available data for $\gamma p \to K^0 \Sigma^+$, $\gamma p \to K^+ \Sigma^0$, $\gamma n \to K^0 \Sigma^0$, and $\gamma n \to K^+ \Sigma^-$ will provide more constraints on theoretical models, especially, in help disentangling the hadronic and electromagnetic resonance couplings. Further work in this direction is under consideration and will be presented elsewhere. This will help prepare for a more complete analysis within a dynamical coupled channels approach to better understand the reaction dynamics of the $K\Sigma$ photoproduction reactions and to extract the resonance parameters from the poles of the reaction amplitude in the complex energy plane.

\begin{acknowledgments}
This work is partially supported by the National Natural Science Foundation of China under Grants No.~12175240, No.~12147153, and No.~11635009, the Fundamental Research Funds for the Central Universities, and the China Postdoctoral Science Foundation under Grants No.~2021M693141 and No.~2021M693142. 
\end{acknowledgments}

\end{document}